# Modeling Spontaneous Exit Choices in Intercity Expressway Traffic with Quantum Walk

*Zhaoyuan Yu, Xinxin Zhou, Xu Hu, Wen Luo, Linwang Yuan, and A-Xing Zhu*

**Abstract**—**In intercity expressway traffic, a driver frequently makes decisions to adjust driving behavior according to time, location and traffic conditions, which further affects when and where the driver will leave away from the expressway traffic. Spontaneous exit choices by drivers are hard to observe and thus it is a challenge to model intercity expressway traffic sufficiently. In this paper, we developed a Spontaneous Quantum Traffic Model (SQTM), which models the stochastic traffic fluctuation caused by spontaneous exit choices and the residual regularity fluctuation with Quantum Walk and Autoregressive Moving Average model (ARMA), respectively. SQTM considers the spontaneous exit choice of a driver as a quantum stochastic process with a dynamical probability function varies according to time, location and traffic conditions. A quantum walk is applied to update the probability function, which simulates when and where a driver will leave the traffic affected by spontaneous exit choices. We validate our model with hourly traffic data from 7 exits from the Nanjing-Changzhou expressway in Eastern China. For the 7 exits, the coefficients of determination of SQTM ranged from 0.5 to 0.85. Compared with classical random walk and ARMA model, the coefficients of determination were increased by 21.28% to 104.98%, and relative mean square error decreased by 11.61% to 32.92%. We conclude that SQTM provides new potential for modeling traffic dynamics with consideration of unobservable spontaneous driver's decision-making.**

## I. INTRODUCTION

The confined nature of intercity expressways with its alternative exists to nearby destinations (local towns and cities) along it creates a unique traveling behavior of drivers. They cannot exit the expressways until they reach the designated exists to their respective destinations. They frequently makes decisions to adjust his/her drive behavior according to time, location and traffic conditions [1]. These choices for multiple exits for a destination often make drivers choose the particular exit rather spontaneously. The aggregation of the spontaneous exit choices by individual drivers will affect the distributions of traffic flow in space-time along the expressway [2]. Besides, due to the dense traffic flow and the relatively fast speed of the expressway traffic, any vehicle that passes through the expressway exit will affect the speed of the following vehicles. The combined effect of a large number of individual spontaneous decisions at different destinations/exits along the expressway makes the overall traffic flow exhibit significant inhomogeneities, nonlinearities, and uncertainties in space-time [3]. Considering this spontaneous behavior of drivers in traffic flow simulation might help to reduce model uncertainty and improve model accuracy.

Current models try to capture the influences of spontaneous exit choices through two major perspectives: deterministic and stochastic, thus lead to two major categories of models: deterministic and stochastic models [4]. The deterministic models consider that spontaneous exit choices can be accurately modeled and solved by numerical methods, and the output of the model is fully determined by the parameter values and the initial conditions. Deterministic models can be further classified into three subcategories: macroscopic models, microphysical models, and intelligent agent models [5]. Macroscopic models consider the overall effects of spontaneous exit choices. Such models commonly use endogenous or exogenous variables to summarize the effects of exit choices (e.g. origin-destination models [6], logit analysis [7]). Therefore, it only reflects the total or average effects of exit choices rather than the spontaneous exit choices of individual drivers [8]. Microphysical models (e.g. gravity models [9], lagrangian traffic state models [10], Lighthill-Whitham and Richards models [11]) use physics laws to approximately describe the influences of spontaneous exit choices. Such models abstract each driver into indistinguishable particles, characterizing the influences of spontaneous exit choices by varying the attractiveness of the particles to approximate the spatial and temporal dynamics of each driver through different exits [12]. Microphysical models can reflect spatial-time distribution of spontaneous exit choices through the change of the value of exit attractions. However, the difference of exit attractions can only be predetermined as the initial condition of the model, thus cannot show how the drivers themselves spontaneously adjust their driving behaviors. The intelligent agent models simulate the spontaneous exit choices through an agent with active sensing and decision-making ability [13]. Such models can support the spontaneous exit choices of drivers by giving the agent

This work was supported in part by the National Natural Science Foundation of China under Grant 41971404, 41625004 and the Ministry of Science and Technology under Grant 2016YFB0502301.

Zhaoyuan Yu, Xinxin Zhou, Xu Hu, Wen Luo, and Linwang Yuan are with the Key Laboratory of VGE (Ministry of Education), Nanjing Normal University, No.1 Wenyuan Road, Nanjing 210023, China(e-mail: yuzhaoyuan@njnu.edu.cn; 181301027@stu.njnu.edu.cn; 191302068@stu.njnu.edu.cn; 09415@njnu.edu.cn; 09368@njnu.edu.cn).

A-Xing Zhu is with University of Wisconsin-Madison, Madison, WI, 53705, USA(e-mail: azhu@wisc.edu).

predefined evolution rules or targets to evolve the agent autonomously [14]-[15]. However, the predefined rules can only be used to model exits choices from the perspective of certainty rather than stochastic [16]. Therefore, the intelligent agent models cannot reveal the impact of dynamic nature of driver's spontaneous exit choices in driving. These deterministic models have the advantages of having clear mechanisms and can be efficiently solved numerically, but they do not have the ability to model the dynamic effects of the spontaneous nature of drivers.

The second category is the stochastic models. Such models assume that the spontaneous exit choices are a complex stochastic system, formed spontaneously under the influence of many factors, and cannot be accurately modeled with simple deterministic physics or evolutionary rules. Stochastic models formulate spontaneous exit choices within a complex network, which considers each choice the driver made as a node in the network. Stochastic models commonly use a classical random walk to generate random paths that transit from one node to others to simulate the dynamical statistical characteristics of the drivers' spontaneous exit choice [17]. The classical random walk assumes the transition probability between different nodes follows a particular probability distribution, thus the dynamics of the whole traffic can be evaluated according to the statistical characteristics of the walk paths on the network [18]-[20]. Recently，with the continuous accumulation and refinement of traffic observation data [14], [15], [21], [22], growing evidences show that the classical random walk can well capture some nonlinear and nonstationary characteristics, e.g. power-law distribution [23], dynamical rerouting behavior [24] and percolation pattern [25], of urban traffics. However, rare study reports good results on applying classical random walk on expressway traffic. This is because in classical random walk, whether each driver leaving the expressway is independent. However, in expressway traffic flow, the driver's departure from the traffic flow during the exit selection process causes the state of the traffic flow to change, thereby affecting the spontaneous exit choices of the drivers in the subsequent traffic flow[26], [27].

There are several key challenges for the current models to model spontaneous exit choices. Firstly, the dynamic decisions of the drivers during the driving process can hardly be observed. Even if the trajectory of a driver can be logged with video or other methods, the detailed mechanisms of how a driver makes spontaneous exit choice is unknown [28]. With this unobservable nature of spontaneous exit choices, the deterministic model can hardly be used. Secondly, spontaneous exit choice is a dynamic process. The probability of when and where a driver will exit the traffic through a particular exit at a specific time varies continuously. The dynamic evolution of spontaneous exit choices in both time and space makes the decision-making space for drivers rather huge. Thirdly, the dynamic process of the driver's spontaneous exit choice is generally in parallel rather than sequentially. In a single spontaneous exit choice, the driver at a given time may consider exiting the expressway at multiple exits which can occur at different times [23]. With this simultaneous decision-making, the process of spontaneous exit choice for multiple alternatives should be modeled by a process of simultaneous search and traversal of multiple alternative paths rather than a single path in the entire decision network. Lastly, there are bidirectional interactions between the spontaneous exit choices and traffic volumes. The aggregation of spontaneous exit choices by many drivers will change the traffic condition, yet, the change of traffic condition will also affect the spontaneous exit choices of each driver. As a continuous dynamic observation of the driver's choices of exits is not possible, the mechanisms of this bidirectional driver-to-traffic interactions are not clear. From the above discussion it is clear that modeling the effects of the dynamic nature of spontaneous exit choices by drivers on traffic conditions is still a challenge yet to be address adequately.

Quantum Walk, is the quantum extension of classical random walk, which is an ideal tool to solve dynamical and parallel stochastic path searching on complex networks [29], [30]. In quantum walk, when the system is not observed, the driver can choose any of the feasible nodes(exits) at the same time with different probabilities. The transition between the nodes on the network is not "either here or there" but "both here and there", thus the path generation in the quantum walk is "all possible paths" rather than "a single path" as in classical random walk [31]. All possible paths then will collapse to a single determined path when the observation is applied. With efficient matrix algorithms [32], a quantum walk can converge much faster than the classical random walk. In some applications, there is an exponential algorithmic speedup than classical random walk [33]. Due to the advantages of the quantum walk, there are already many applications of quantum walk in algorithm construction [34], decision-making simulation [35] and data analysis [36].

In this paper, we try to overcome the above challenging of modeling and simulation of spontaneous exit choice by introducing the quantum walk models. By integrating the spontaneous exit choices from an interactive quantum probability perspective, we developed the Spontaneous Quantum Traffic Model (SQTM) to simulate the stochastic traffic fluctuation caused by spontaneous exit choices by a quantum walk (QW) and the residual regularity fluctuation by autoregressive moving average (ARMA) model. Our paper is organized as follows: Section II formulates the research problems. Section III introduces the overall modeling process. Section IV presents a case study using Nanjing-Changzhou Expressway with 7 toll stations serving as a validation. Discussion is provided in Section V and conclusions are drawn in Section VI.

## II. PROBLEM DEFINITION AND BASIC IDEA

### A. Formal Definition of the Problem

Consider there is an expressway with $n$ different exits, there are totally $m$ drivers on the expressway at a specific time $T_k$. For each exit $E_i$, there is an observation station (e.g. toll station or video surveillance camera) logs $F_{ik}$, which is the total number of vehicles leave the expressway through the exit $E_i$ at time $T_k$ with a sampling time interval $\Delta T_k$. Generally, we assume all the $i, j, k$, which are the indexes of exit, driver and time begin from 1, and $i = \{1, 2, \cdots, n\}$, $j = \{1, 2, \cdots, m\}$ and $k = \{1, 2, \cdots, s\}$. In the real world, the only observable variable is the $F_{ik}$, which is also one of the most important factors to indicate the characteristics of the expressway conditions. Obviously, $F_{ik}$ is continuously changing over time, and we hope to simulate the dynamics of $F_{ik}$ as preciously as possible.

We assume that $F_{ik}$ consists of two fundamental parts: the deterministic part $F_{ikd}$ and the random part $F_{ikr}$. We assume that the $F_{ikd}$ is a combination of completely predictable components, such as the mean value and the regular cycles of the traffic flow. Then the $F_{ikr}$ is the unobservable random variation caused by spontaneous exit choice. So, we have the fundamental equation of traffic flow in each exit $E_i$ is:

$$F_{ik} = F_{ikd} + F_{ikr} \qquad (1)$$

As the $F_{ikd}$ is totally predictable, which means we can use a simple model, like autoregressive moving average model [37], to approximate it. The major issue is how to simulate the dynamics of spontaneous exit choices to improve the predictability of the $F_{ikr}$.

There are three basic procedures that are required to simulate $F_{ikr}$ by considering spontaneous exit choices. The first is how to model all possible choices for a spontaneous exit. The second is how to represent the dynamical evolution of spontaneous exit choices. The third is how to construct the relationship between the dynamics of spontaneous exit choices and $F_{ikr}$. In the following sections, we will present our solution in modeling $F_{ikr}$ in three procedures. We first formulate the problem associated with each of these procedures and then describe our basic idea in addressing these problems.

**Procedure 1:** Modeling all the possible choice making of a spontaneous exit. The essence of spontaneous exit choice is a driver chooses an exit to leave the expressway traffic at a specific time. Therefore, the key representation of the choice for a spontaneous exit is the presentation of a driver choosing which exit at what time with how much probability. Assuming that the decision process of any driver $D_j$ expected to exit the highway from at exit $E_i$ at time $T_k$ is a node $V_{(i,k)}$, and the complete set of exit choices is $E_i \times T_k$. Therefore, in time and space, all possible spontaneous exit choices of any driver $D_j$ can be combined into a tree-style network. Any node $V_{(i,k)}$ on the network is a state that a chosen spontaneous exit, and any edge $E$ represents the spatiotemporal evolution of a spontaneous exit choice of a given driver. As both the traffic direction and time of expressway traffic is in one direction, the path on the network can only be generated sequentially, therefore, we can reformulate the $n \times s$ network into a tree-style

network as is shown in Fig. 1. We named this tree-style network as the spontaneous exit choice decision space. For each node of this decision space, a probability vector can be assigned to express the driver's preference for the decision. Generally, a n-dimensional probability vector $P(i, j, k) = (p_1, p_2, \cdots, p_n)$, where $(p_1, p_2, \cdots, p_n)$ can be used to indicate the probability of a driver at any time $T_k$ to choose exit $(E_1, E_2, \cdots, E_i)$ to exit the traffic. Clearly, the probability vector $P(i, j, k)$ is unobservable.

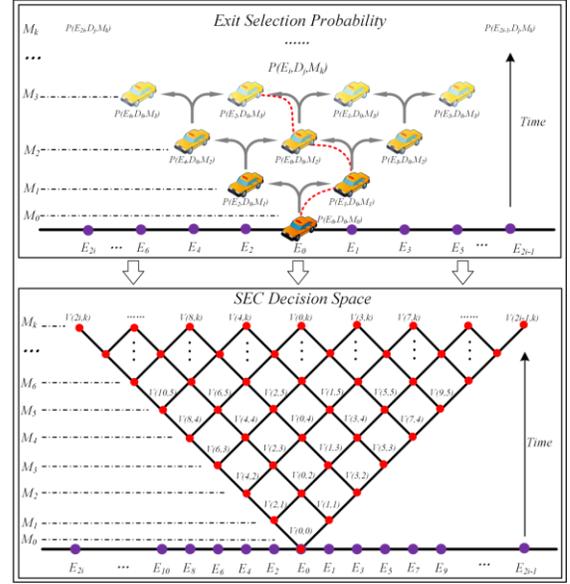

Fig. 1. The quantum walk expressway traffic simulation model

**Procedure 2:** Modeling the dynamical evolution of spontaneous exit choices. The dynamics of spontaneous exit choices can be formulated as a dynamical path generation according to the dynamic change of the probability vector. Typically, the dynamic change of the probability vector can be expressed by the state transition matrix, namely:

$$P(i, j, k)_{s1} = T_{s0 \to s1} P(i, j, k)_{s0} \qquad (2)$$

The full matrix expression form of (2) is:

$$\begin{pmatrix} p_1'(i, j, k) \\ \vdots \\ p_n'(i, j, k) \end{pmatrix} = \begin{pmatrix} t_{11} & \cdots & t_{1n} \\ \vdots & \ddots & \vdots \\ t_{n1} & \cdots & t_{nn} \end{pmatrix} \bullet \begin{pmatrix} p_1(i, j, k) \\ \vdots \\ p_n(i, j, k) \end{pmatrix} \qquad (3)$$

where $T_{s0 \to s1} = \begin{pmatrix} t_{11} & \cdots & t_{1n} \\ \vdots & \ddots & \vdots \\ t_{n1} & \cdots & t_{nn} \end{pmatrix}$, is the state transition matrix, which reflects the transition process of spontaneous exit choices of a driver between different states.

In the real traffic flow process, not only the probability vector $P(i, j, k)_{s0}$ and $P(i, j, k)_{s1}$ but also the transition matrix $T_{s0 \to s1}$ are unobservable. As there are heterogonies in exit selection between different drivers, each driver has a specific transition matrix $T_{s0 \to s1}$.

In real-world spontaneous exit selection, a driver indeed considers all the feasible choice as a total, which means in the network, he/she chooses every possible node available to generate multiple paths. These paths evolve through time and the probability distribution of its position at any node is not deterministic. In addition, any changes in the selection of the probability of a node will affect all the probability distribution. As the probability is unobservable, the continuous change of probability will, in turn, create a chain reaction, which will then lead to a combinatorial explosion. This combinatorial explosion makes it difficult, if not impossible, to use deterministic or stochastic methods for both modeling and simulating the spontaneous exit choices of massive drivers in expressway traffic.

**Procedure 3:** Mapping the dynamics of spontaneous exit choices into real traffic volume. In order to simulate the real volume of traffic flow at each exit of the expressway, it is necessary to resolve the mapping between spontaneous exit choices of drivers and the dynamic of traffic flow at each exit. Since the simulation of spontaneous exit choices only output the probability of whether a driver will choose a certain exit, a calibration is needed to map this probability to the real volume of the traffic flow. Since the observation of the state of each vehicle on the highway is not practical, the calibration and verification of the model can only be operated on the total amount of traffic observed at the exit.

A generalized calibration model can be constructed as follows:

$$F_{ikr} = f\left(P\left(i,k\right)\right) \qquad (4)$$

Where $F_{ikr}$ is the real traffic volume obtained by simulated probability, $P\left(i,k\right)$ is the simulated probability of leaving the expressway from exit $E_i$ at time $T_k$, and $f$ is the calibration function that maps $P\left(i,k\right)$ to $F_{ikr}$.

### B. Basic Idea

The unobservable and combinatorial explosion nature of spontaneous exit choices, which are key difficulties to model under the classical statistical framework, can be modeled in the quantum walk. Different from a classical random walk, which uses a probability vector that forces the driver to choose any node in the decision space of spontaneous exit choices, quantum walk assumes that at any time the driver will select all the possible choice, i.e. nodes, at a single time. Quantum walk uses state vector, which is a vector that expresses how relatively a driver will choose each node at a time, to represent the decision making. With the state vector, a dynamical probability distribution with each node to show the preference of the driver can be generated. As the probability distribution of state vector is unobservable at any time, quantum walk chooses several decisions in parallel with the linear combination of the state vector (See details of the superposition in Section III. A.). It is more realistic in spontaneous exit choice selection that any driver will make a decision in parallel and select more than one choice at a single time. So, with the state vector representation, we can solve the unobservable probability problem in Procedure 1.

In quantum walk, the characteristics of the dynamic path generation is determined by a unitary process rather than a stochastic process. The unitary process, which is similar to a coefficient matrix $A$ in the linear equation $Y = AX$, means that it is deterministic and completely reversible. Therefore, even if we do not know the exact form of $T_{s0 \to s1}$ in (2), we can analytically evaluate the equations without know the actual value of $T_{s0 \to s1}$. Thus, the quantum walk can still generate "all possible paths" without any prior assumption of probability distribution of each node. Although there's seems to be a combinational explosion if we consider "all possible paths" all the time during the evolution, the "quantum collapse", which happened when the observation is applied, will map all the possible paths into a single determined path. We can then solve the combinatorial explosion problem in Procedure 2. With the solution of both Procedure 1 and Procedure 2, we can develop simple calibration method to map the state vector to the traffic volumes as shown in Procedure 3.

With the above discussion, we introduce quantum walk to simulate the random traffic fluctuation caused by drivers' spontaneous exit choices. By developing the quantum state vector representation of the driver's preference of choices, the quantum walk can be applied to simulate the dynamic updating process of spontaneous exit choices. With the relationship between quantum state vector and the probability distribution for each choice, the probability of when a driver will leave the expressway through which exit can be generated. With the calibration of this simulated probability distribution to real observational traffic volume data, the traffic fluctuation caused by the driver's spontaneous exit choice can be generated. The integration of the deterministic part $F_{ikd}$ with the random part $F_{ikr}$ formulates our unified model, the Spontaneous Quantum Traffic Model (SQTM). The overall framework of SQTM is shown in Fig. 2.

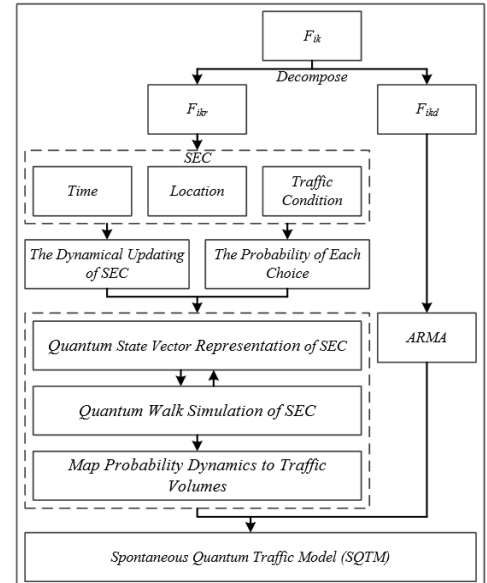

Fig. 2. The overall framework

### III.  The Sqtm Model

#### A.  *Quantum State Vector Representation of Spontaneous Exit Choices*

Suppose $G = (V, E)$ is the road network graph without considering its width, where $V$ is the set of nodes, which represents the possible choices at each exit, $E$ is the set of edges, representing the possible evolution path of spontaneous exit choices over time. The number of nodes and edges in G are $|V|$ and $|E|$, respectively. For any node $v$, $\Gamma_v = \{u \in V : (u, v) \in E\}$ represents the node adjacent to $v$ (neighbor node). $A$ is the adjacency matrix of network $G$, that is:

$$A_{uv} = \begin{cases} 1, if (u, v) \in E \\ 0, otherwise \end{cases} \tag{5}$$

$G$ is a network graph without width, directions and loops, so $A_{uv} = A_{vu}$ and $A_{vv} = 0$. Since different exits are connected by the expressway, the moving of drivers in the expressway can be modeled as transitions among different exits. At any time, a binary state indicating whether a driver will or will not exit the expressway (i.e. select or not select a node) can be constructed for each node. At any moment, the driver will choose a strategy to adjust the driving behavior. Therefore, all the selected strategy states of a driver can be defined as the base state of quantum walk, which is recorded as $|v\rangle$. For each decision, all nodes are mutually exclusive, which means that all base states are orthogonal to each other. Therefore, these base states can be expressed in the form of vectors, as follows:

$$|1\rangle = \begin{pmatrix} 1 \\ 0 \\ \vdots \\ 0 \end{pmatrix}, |2\rangle = \begin{pmatrix} 0 \\ 1 \\ \vdots \\ 0 \end{pmatrix}, \cdots, |i\rangle = \begin{pmatrix} 0 \\ 0 \\ \vdots \\ 1 \end{pmatrix} \tag{6}$$

The dynamic exit selection made by different drivers can be seen as a random process and thus can be modeled with a bidimensional variable with another two states at each node as "exit $|a\rangle$" and "not exit $|b\rangle$", respectively. Since drivers cannot have both states of $|a\rangle$ and $|b\rangle$ at the same time, the two states are orthogonal (Fig. 3). However, given that the state vector of every driver's changes during a time period, there will be a probability distributions of state changes of both $|a\rangle$ and $|b\rangle$. Therefore, we can formulate a vector of a complex number to express the state of any driver:

$$\varphi_k = a_k |a\rangle + b_k |b\rangle i \tag{7}$$

We call this $\varphi_k$ as a superposition state. $|a_k\rangle$ and $|b_k\rangle$, which fits the relationship that $|a_k|^2 + |b_k|^2 = 1$ is called the probability amplitudes of states $|a\rangle$ and $|b\rangle$. Thus $|a_k|^2$ and $|b_k|^2$ are identical to the probability of a driver at states $|a\rangle$ and

$|b\rangle$ in the whole $\Delta k$ time period, respectively. This complex number will exist if we do not observe the driver during the whole driving process. If we observe the driver at any time, the state vector of the driver will allocate into the binary state $|a\rangle$ or $|b\rangle$.

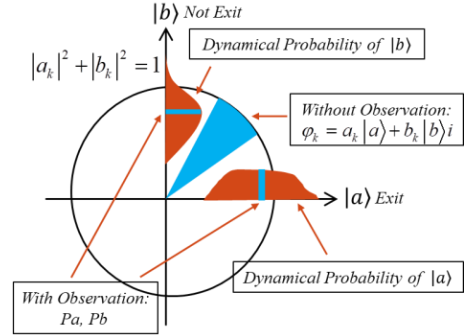

Fig. 3.  The geometric representation of quantum state vector

The quantum state vector representation, which considers the stochastic process from a dynamic perspective, provides an ability to describe an individual driver's decision of exit selection. Any decision made by an individual driver affects the state vector trajectory on the unit cycle, therefore the decision can only be modeled or investigated as stochastic processes. This geometric representation, which can be seen as a generalized representation of probability, can be used to model the overall dynamics of the traffic flow. Since the model will converge into definite states when the observation is applied, it also possible to use observation data as a constraint to solve the model, which may greatly reduce the complexity of the model.

#### B.  *Quantum Walk Simulation of Spontaneous Exit Choices*

Before we formulate the quantum walk, we first define a Hilbert space $H \in C^N$, which is made up of base state $|v\rangle$. The characteristic of Hilbert space $H$ is that any quantum superposition state is a linear combination of base states, and the superposition state is also in $H$. Therefore, the state $|\varphi(k)\rangle$ of any driver with a quantum walk on $G$ can be defined as a linear superposition state of all base states, which is expressed as follows:

$$|\varphi(k)\rangle = \sum_{v \in V} a_v(k) |v\rangle \tag{8}$$

Where $a_v(k)$ is the probability amplitude of the corresponding base state $|v\rangle$ at time $T_k$, $a_v(k) \in C$ and $|a_v(k)| \in [0, 1]$. The probability of finding the walker at any exit at time $T_k$ is given by the squared modulus of the appropriate element of $\varphi(k)$ [33]. That is, the product of the probability amplitude and its complex conjugate is equivalent to the probability of the walker appearing on a node in the classical space. Therefore, the probability that a driver is in base state $|v\rangle$ at time $T_k$ can be expressed as:

$$p\left(\left|v\right\rangle,k\right)=\alpha_v(k)\alpha_v^*(k) \qquad (9)$$

Where $\alpha_v^*(k)$ is a complex conjugate of $\alpha_v(k)$. At time $T_k$, it satisfies $\sum_{v\in V}p\left(\left|v\right\rangle,k\right)=1$.

Unlike the classical random walk, the process of quantum walk is not a Markov chain [33]. The evolution of state vector $\left|\varphi(k)\right\rangle$ with time $T_k$ is realized by the following unitary transformation:

$$\frac{d}{dt}\left|\varphi(k)\right\rangle=iA\left|\varphi(k)\right\rangle \qquad (10)$$

It can be seen from (10) that the state evolution process depends on the adjacency matrix $A$. However, as a differential equation with complex coefficients, it is difficult to solve it directly. Therefore, it can be transformed into a matrix form from the discrete point of view, and the quantum walk can be efficiently solved by matrix operation.

Assuming a driver starts at the node $\left|v\right\rangle$, from (9), we know that it will have a state vector:

$$\left|\varphi(k)\right\rangle=e^{-iA}\left|v\right\rangle \qquad (11)$$

Then the probability $p_{vu}(k)$, indicating the walker moves from node $v$ to node $u$, is:

$$p_{vu}(k)=\left\|\mu\left|\varphi(k)\right\rangle\right\|^2 \qquad (12)$$

Analogously to (3), we can express the state vector as a matrix:

$$\begin{pmatrix}\varphi_1'\\\vdots\\\varphi_n'\end{pmatrix}=\begin{pmatrix}\mu_{11}&\cdots&\mu_{1n}\\\vdots&\ddots&\vdots\\\mu_{n1}&\cdots&\mu_{nn}\end{pmatrix}\bullet\begin{pmatrix}\varphi_1\\\vdots\\\varphi_1\end{pmatrix} \qquad (13)$$

Where $\mu$ is a unitary matrix varying with time. Similar to (11) and (13), we can construct a time evolution operator $U(k)=e^{-iAk}$ to construct the dynamical evolution quantum walk [31]. As the topology of the expressway is a line, we have:

$$A=\begin{bmatrix}0&1&0&\dots&0&0\\1&0&0&\dots&0&0\\0&1&0&\dots&0&0\\\vdots&\vdots&\vdots&\ddots&\vdots&\vdots\\0&0&0&\cdots&1&0\\0&0&0&\dots&0&1\\0&0&0&\cdots&0&0\end{bmatrix} \qquad (14)$$

As $A$ is a matrix and the numerical solution of such a model is complex, we can use a polynomial expansion to approximate the quantum walk [32]. From (11), we have:

$$e^{-iHk}=\sum_{n=0}^{N-1}c_nA^n \qquad (15)$$

Where $N$ is the number of distinct eigenvalues of $A$, and $c_n$ are unknown coefficients that must be determined. These coefficients can be fixed by using the (15) and continue to be

valid when $A$ is replaced by each of its eigenvalues. Assuming the Tylor expansion of $U(k)$ as:

$$U(k)=c_0I+c_1A+c_2A^2+c_3A^3+\cdots+c_nA^n \qquad (16)$$

Where $A$, which can be seen as $A^0$, is the identity matrix, and the $c_n$ is the weight coefficients to be determined. Our evaluation of the time evolution operator is based on the Cayley-Hamilton theorem [38], which states that every square matrix satisfies its own characteristic equation:

$$det(A-I\lambda)=0 \qquad (17)$$

Where $A$ is the original matrix, $I$ is the identity matrix, and $\lambda$ is the eigenvalue. The characteristic equation is a polynomial equation in $\lambda$ and will remain valid when $\lambda$ is replaced by $A$. With the Cayley-Hamilton theorem, we can replace the Hamiltonian with each of its eigenvalues in the previous equation to get the following system of equations:

$$\begin{bmatrix}e^{-i\lambda_1k}\\e^{-i\lambda_2k}\\\vdots\\e^{-i\lambda_{n-1}k}\\e^{-i\lambda_nk}\end{bmatrix}=\begin{bmatrix}1&\lambda_1&\lambda_1^2&\cdots&\lambda_1^{n-1}\\1&\lambda_2&\lambda_2^2&\cdots&\lambda_2^{n-1}\\\vdots&\vdots&\ddots&\cdots&\vdots\\1&\lambda_{n-1}&\lambda_{n-1}^2&\cdots&\lambda_{n-1}^{n-1}\\1&\lambda_n&\lambda_n^2&\cdots&\lambda_n^{n-1}\end{bmatrix}\begin{bmatrix}c_1\\c_2\\\vdots\\c_{n-1}\\c_n\end{bmatrix} \qquad (18)$$

One can solve for the coefficients with simple linear algebra to obtain an expression for the time evolution operator: a $n\times n$ matrix. With this matrix solution, we can then solve the (12) to get the quantum walk simulation. With (18), we can have the dynamical change of probability of each node. Therefore, the dynamics of spontaneous exit choices can be simulated.

## C. Map Probability Dynamics to Traffic Volumes

In the quantum walk-based expressway simulation model, there are two key parameters that need to be determined. One is the time step $\Delta k$ of the quantum walk evolution, and the other is the correction factor $a_i$ from the probability magnitude to the real traffic data. Assuming that the time series of the traffic flow of the exit obtained by the real observation is $F_{ik}$, and the quantum walk of the parameter is $\Delta k$, the $i^{\text{th}}$-exit probability sequence of the model output is $P_i(k)$, then the quantum walk correction model can be constructed as follows:

$$F_{ik}=a_iP_i(k) \qquad (19)$$

In the above model, $P_i(k)$ is a probability determined by the characteristics of quantum walk. The variable $k$ determines the characteristics and laws of the quantum walk probability evolution, which determines the time scale and period characteristics of the quantum walk simulation data (see the Discussions section). The parameterization is carried out by means of stepwise approximation. Referring to the determination method of the link coefficient in the classical random walk, we can gradually increase $k$ with an interval of $\Delta k$, and compare different simulation results with the measured data in the overall waveform structure and statistics. The best match will produce the optimal $\Delta k$. After the optimal $\Delta k$ parameter is determined, the correction coefficient $a_i$ can be obtained by directly fitting the simulation results to the real observation using the least square method.

As there is a large amount of regular traffic that periodically changes in the traffic flow, which cannot be captured by the quantum walk, we also use the autoregressive moving average model to model the residual series.

The autoregressive moving average model is composed of autoregressive model and moving average model, which is an important method to study time series, and the corresponding parameters are $p$ and $q$ respectively. The $p$ denotes the lag number of the time series data used in the prediction model, expressed as the partial autocorrelation coefficient which is close to 0 in the $p$-order lag, also called AR/auto regressive term. The $q$ represents the prediction used in the moving average prediction model. The hysteresis of the error is expressed as the autocorrelation coefficient approaching 0 at the $q$-order lag, also known as the MA/moving average term. The overall formula of autoregressive moving average model is:

$$y_t = \mu + \phi_1 \times y_{k-1} + \cdots + \phi_p \times y_{k-p} + \theta_1 \times \varepsilon_{k-1} + \cdots + \theta_p \times \varepsilon_{k-p} \quad (20)$$

where $\phi$ is the coefficient of the auto regressive process and $\theta$ is the coefficient of the moving average process.

To summarize, the overall model of SQTM is:

$$SQTM(F_{ik}) = QW(k) + ARMA(p,q) = \alpha_i P(i,k) + ARMA(p,q) \quad (21)$$

Where, $i,k$ is the index of exit and time respectively, $\alpha_i$ is the correction factor from the probability $P(i,k)$ to the real traffic data, $p$ and $q$ is the fitting parameter of ARMA.

## IV. EXPERIMENTS

### A. Research Data and Experiment Configuration

Shanghai-Nanjing Expressway has one of the busiest traffic flows with complicated conditions in eastern China [39]. This paper selected the Nanjing-Changzhou section of the Shanghai-Nanjing Expressway as the study area and selected the distribution of vehicles departing from Nanjing and passing through 7 exits(Fig. 4) during the period from Dec-01-2015 to Dec-30-2015. To form the intercity expressway traffic, we only selected the vehicles entering directly from Nanjing at each exit. The original data contains information about each vehicle passing through the exit, so the volume is huge. To reduce the data volume, we accumulate the number of vehicles that pass each exit in one hour. Then, we have the hourly time series for each exit. The overall sample points for each exit are 31(days)×24(hours)=744 (time points).

Based on the distribution of each exit in the original data and the road topology, the adjacency matrix between each exit was constructed. The quantum walk simulation experiment was conducted based on the adjacency matrix. The control parameters of the quantum walk mainly are $\Delta k$ and $\alpha_i$, which determines the time interval of the simulated unit time through the exit of the vehicle, thus affecting the time scale characteristics of the simulated time series (length and number of cycles, etc.) over a long time period. To this end, we traverse different $\Delta k$ ways to obtain the value of the optimal control parameter $\Delta k$ (for detailed analysis, see Discussion). For this, $\Delta k$ was gradually increased from 0 to 1 at the interval of 0.01, by comparing the statistics of different $\Delta k$ values, the best value of $\Delta k$ can be determined. The probability sequence obtained by the simulation is subjected to calibration, which is mapped into the total sequence of real observation data, the model performance and statistics were calculated, and the residual analysis was performed. To further demonstrate the performance of the quantum walk, the classical random walk method was used at the same time for comparison. The connection degree of the classical random walk is also determined by the traverse method. In order to further reveal the regularity of the residual part, the autoregressive moving average model is used to fit the residual sequence. Finally, the whole experiment is completed to verify the SQTM.

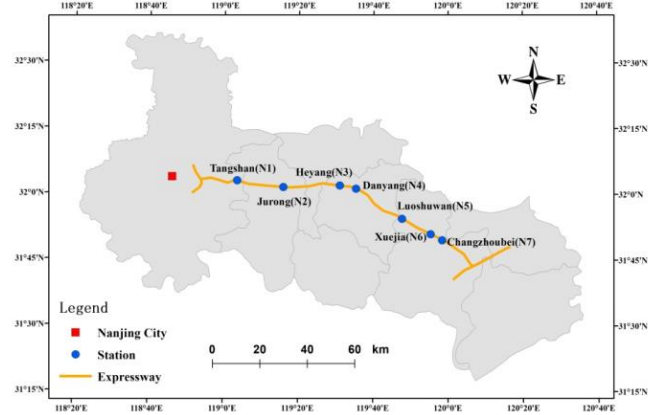

Fig. 4. Research area and exits distribution

For the performance of SQTM, this paper used three statistical indicators to evaluate it, including the mean absolute error(MAE), the root mean square error(RMSE) and the coefficient of determination($R^2$). The definition of these statistical indicators is listed in Table I. In Table I, $y_i$ is the real observation data, $\overline{y_i}$ is the average of the real observation data, and $y_i$ is the modeling data.

TABLE I
Model performance evaluation indicators

| Statistical Index | Definition |
|---|---|
| Mean Absolute Error(MAE) | $MAE = \dfrac{1}{m}\sum_{i=1}^{m}\lvert y_i - y_i \rvert$ |
| Root Mean Square Error(RMSE) | $RMSE = \sqrt{\dfrac{1}{m}\sum_{i=1}^{m}\left(y_i - y_i\right)^2}$ |
| Coefficient of Determination(R2) | $R^2 = 1 - \dfrac{\sum_{i=1}^{m}\lvert y_i - y_i \rvert^2}{\sum_{i=1}^{m}\lvert y_i - \overline{y_i} \rvert^2}$ |

## B. Results

Based on the experiment configuration, the average statistics of quantum walk simulation with different $\Delta k$ values were listed in Table II, when $\Delta k = 0.13$, the statistics were the best, so the optimal fitting parameter $\Delta k$ of quantum walk for the 7 exits was set to 0.13. Besides, the coefficients of SQTM were listed in Table III. The $\alpha_i$ coefficient for 7 exits was between 25.23 and 503.42, which is proportional to the total number of vehicles passing through each exit. For coefficients of the autoregressive moving average model for the regularity part $F_{ikd}$, it is obvious that the autoregressive coefficient of quantum walk residual series modeling has same the coefficients of $p = 2$, except N3 with which $p = 3$. This indicates that a significant time correlation of 2 hours in the fluctuation structure of vehicle flow series which is consistent with the situation that the travel time from Nanjing is less than 2 hours in drive to Changzhou. From the coefficient of moving average coefficient, the moving average coefficients $q$ of quantum walk modeling residuals are larger than the $q$ of classical random walk modeling residuals, which indicating the residual of quantum walk modelling are more stationary than the classical random walk modelling.

TABLE II Average statistics of quantum walk simulation with different $\Delta k$

| $\Delta k$ | MAE | RMSE |
|---|---|---|
| 0.1 | 25.04 | 32.86 |
| 0.13 | 23.62 | 31.28 |
| 0.2 | 25.23 | 32.74 |
| 0.3 | 25.34 | 32.81 |

TABLE III The coefficients of SQTM

| Exits | Methods | $\alpha_i$ | ARMA Model $(p, q)$ |
|---|---|---|---|
| N1 | STQM | 503.42 | ARMA(2,5) |
| | "RW+ARMA" | 279.91 | ARMA(5,2) |
| N2 | STQM | 33.56 | ARMA(2, 2) |
| | "RW+ARMA" | 16.96 | ARMA(3,1) |
| N3 | STQM | 71.19 | ARMA(3, 1) |
| | "RW+ARMA" | 47.75 | ARMA(1,1) |
| N4 | STQM | 80.28 | ARMA(2, 3) |
| | "RW+ARMA" | 69.03 | ARMA(2,2) |
| N5 | STQM | 25.23 | ARMA(2, 3) |
| | "RW+ARMA" | 16.15 | ARMA(2, 3) |
| N6 | STQM | 75.91 | ARMA(2, 3) |
| | "RW+ARMA" | 51.00 | ARMA(5,1) |
| N7 | STQM | 38.81 | ARMA(2, 5) |
| | "RW+ARMA" | 34.85 | ARMA(2,2) |

The simulation results based on the SQTM are shown in Fig. 5, respectively. We can find that SQTM reproduces the real observation data well. The distribution of peaks and troughs obtained in different time periods is consistent with the law of quasi-periodic fluctuation found in traffic origin-destination matrix in recent studies [40], [41]. The statistics of the model performance of STQM and "RW + ARMA" are listed in Table IV, the $R^2$ of SQTM ranged from 0.5(N5) to 0.85(N1).

Compared with "RW + ARMA" model, the increment of $R^2$ is from 21.28%(N7) to 104.98%(N3), the decrement of RMSE is from 11.61%(N2) to 32.92%(N1), and the decrement of MAE is from 13.42%(N2) to 30.34%(N1). In general, SQTM can model the traffic flow of each exit well.

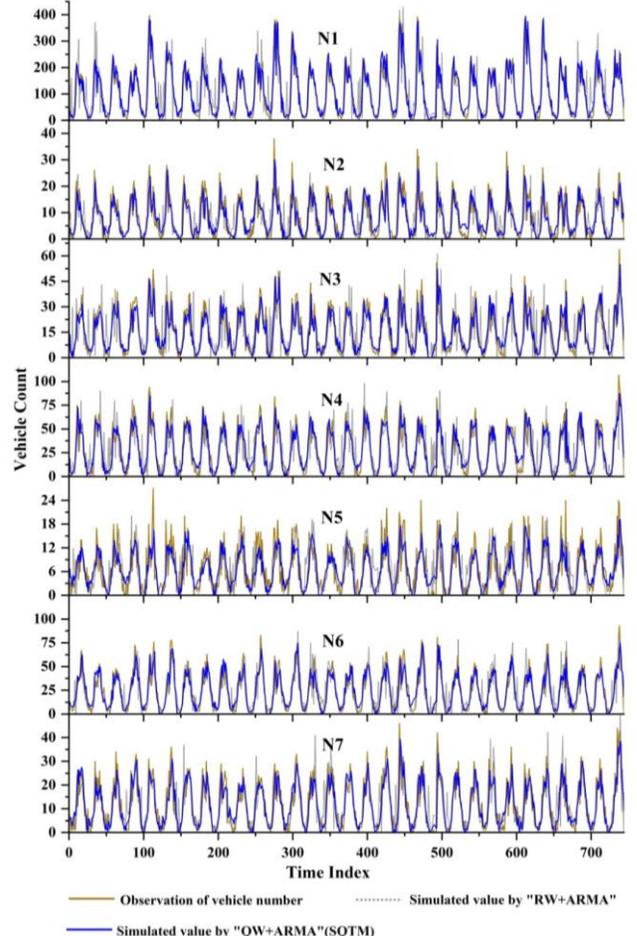

Fig. 5. The simulation results of SQTM

TABLE IV
Statistics of the model performance of STQM and "RW + ARMA"

| Exits | Methods | MAE | RMSE | $R^2$ |
|---|---|---|---|---|
| N1 | STQM | 24.84 | 34.71 | 0.85 |
| | "RW+ARMA" | 35.66 | 89.32 | 0.67 |
| N2 | STQM | 3.33 | 4.76 | 0.57 |
| | "RW+ARMA" | 3.91 | 5.66 | 0.45 |
| N3 | STQM | 5.17 | 7.11 | 0.66 |
| | "RW+ARMA" | 7.39 | 13.44 | 0.32 |
| N4 | STQM | 7.93 | 10.95 | 0.76 |
| | "RW+ARMA" | 11.13 | 17.29 | 0.55 |
| N5 | STQM | 2.80 | 3.77 | 0.50 |
| | "RW+ARMA" | 3.30 | 4.85 | 0.34 |
| N6 | STQM | 7.00 | 9.43 | 0.77 |
| | "RW+ARMA" | 8.95 | 14.50 | 0.61 |
| N7 | STQM | 4.06 | 5.46 | 0.68 |
| | "RW+ARMA" | 4.69 | 9.80 | 0.56 |

## V. Discussion

### A. Improvement of modeling $F_{ikr}$ over classical random walk

In this section, we provide detailed comparison of the modeling of the spontaneous exit choices using quantum walk and that using the classical random walk through three typical exits (N1, N3 and N7). The comparison results of quantum walk and classical random walk are shown in Fig. 6. From Fig.6, it is clear that the amplitude of quantum walk simulation varies in a quasi-period, which fit much better with the observation data. Firstly, there are more high frequency fluctuations in classical random walk simulation results (such as 0 to 50, 125 to 150, and 430 to 470 time intervals of N1 and N3 exits), which have obvious scale differences with the real observation data. However, quantum walk has no such high frequency random fluctuations. Secondly, the distribution of peaks and troughs simulated by quantum walk is relatively uniform without obvious aggregation. For the three exits, the total number of peaks of the actual observation data is 31 peaks for each exit), and the total number of peaks with good fitting performance of quantum walk and classical random walk is 25 and 16, respectively, so the corresponding proportion is 31:25:16. Therefore, the fitting performance of quantum walk on peaks is better than that of classical random walk. Thirdly, because quantum walk only simulates the randomness of the expressway traffic flow, it is normal that there is a certain deviation between the simulated value and the real observation data. However, compared with classical random walk, the simulation results of quantum walk are more consistent with the distribution trend of actual data, which can better recover the change trend of expressway traffic flow. In conclusion, compared with classical random walk, quantum walk can better reveal the randomness of expressway traffic flow.

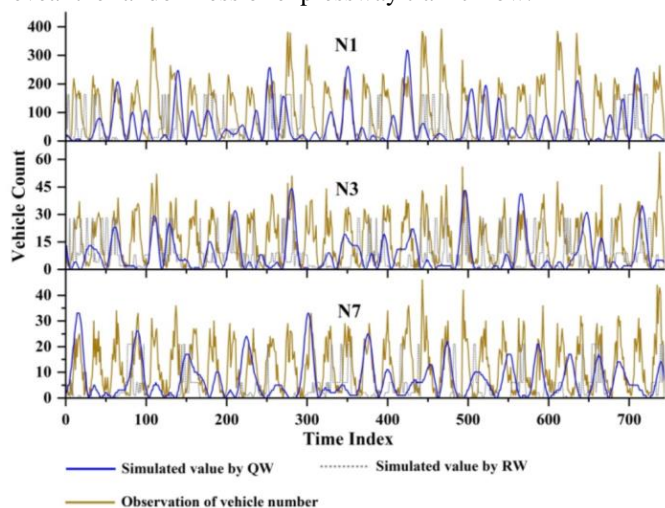

Fig. 6. Comparison between simulation results and real observation data

### B. Applicable issues and future works of quantum walk

Due to the ideal abstraction of the expressway traffic, our SQTM model still has large potential to be improved from the following aspects. Firstly, the research data used in this paper is hour-scale traffic flow observation data. It did not use traffic data at high temporal resolution, such as minute-scale, or even individual vehicle trajectory. The vehicle type was also not used. In future applications, more detailed expressway traffic flow data can be considered which might improve the modeling accuracy. Secondly, we only considered the single lane in SQTM, the lane change situation was not considered. Integrating lane change information may also improve the performance of SQTM. Thirdly, this study did not consider the differences in spatial distance between different exits. The spatial distances between different exits would have an impact on the driver's spontaneous exit choices. In the candidate exits of the driver, most of these exits will not be too far away from their destinations where the spontaneous decision occurs, particularly when the traffic is heavy. In a similar process, the spatial distance between exits will directly affect the probability of the driver driving out of the expressway traffic flow, which needs to be weighted in the quantum walk.

## VI. Conclusion

This paper developed a SQTM model in an effort to model the effects of the spontaneous exit choices of drivers on intercity expressway traffic flow. The SQTM models the deterministic part with an autoregressive moving average model and the randomness caused by the spontaneous exit choices of drivers with quantum walk. For the 7 exits, the coefficients of determination of SQTM ranged from 0.5 to 0.85, with an increment of 21.28% to 104.98% compared with classical random walk and ARMA model. Meanwhile, the RMSE decreased by 11.61% to 32.92%, respectively. SQTM can better reproduce the stochastic fluctuations caused by spontaneous decision-making. Our method provides a good potential for modeling the unobservable driver's spontaneous exit choice on traffic flow simulation.

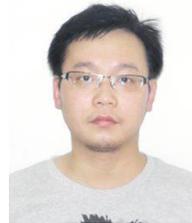

**Zhaoyuan Yu**, , born in Nanling, Anhui in September 1984, PhD, associate professor. In 2005, he graduated from Nanjing Normal University with a major in Geographic Science. After graduation, he was admitted as a graduate student in Nanjing University Normal University majoring in Cartography and Geographic Information System. He is engaged in the research of geographic information systems and geographic modeling and has carried out systematic work in the fields of spatio-temporal data models and methods in the fields of tourism and spatial structures analysis. He has published more than 70 journal papers, directed over 5 national projects and participated in over 20 national projects, including 973 special program, 863 and NSFC Key programs. He was been awarded with 4 prizes including the first prize of natural science of the Ministry of Education, the first prize of national geographic information technology progress, etc.

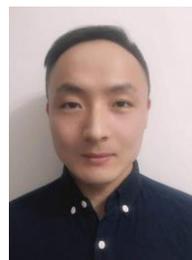

**Xinxin Zhou**, born in Wuhu, Anhui in December 1991. He is currently a PhD candidate in GIS at Nanjing Normal University. He obtained his bachelor's and master's degrees respectively from Nanjing Xiaozhuang University and Nanjing Normal University in GIS. His research interests primarily involve modeling and predicting of human mobility patterns among intercity, learning smart strategies for urban transportation, and traffic flow prediction from fine-grained GPS and sensor network data. Previously, he had worked a long time at the department of TDC (Technology Development Center) of Nanjing Guotu Inc., an award-winning firm that specializes in geospatial technology integration and land planning in China.

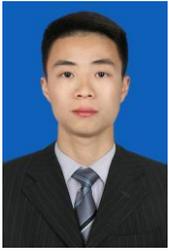

**Xu Hu** received the bachelor's degree in Geographic Information Science from Southwest Petroleum University, Chengdu, China. He is currently pursuing the master's degree with the school of geography and science, Nanjing Normal University, China. He is currently involved in expressway traffic flow simulation, especially on the simulation of expressway traffic flow by quantum random walk.

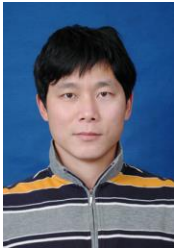

**Wen Luo**, born in Jingzhou, Hubei in September 1987, PhD, associate professor. He graduated from Hubei University in 2008 with a master's degree, and he received his master's and doctor's degrees in Cartography and Geographic Information System from Nanjing Normal University in 2011 and 2014 respectively. He is engaged in the research of algorithms and applications of geographic information systems. In recent years, he has published more than 40 academic papers and 37 SCI and EI journal papers, of which 12 are the first author and corresponding author. He is currently chairing 2 NSFC Projects, and has participated in 1 National Natural Science Foundation key project and three National Natural Science Foundation general projects. Apply for 8 software copyrights and patents.

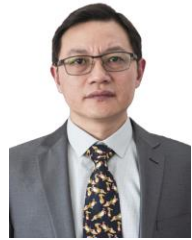

**Linwang Yuan**, born in Hai'an, Jiangsu Province in November 1973, is now a professor in the school of geography and science of Nanjing Normal University. He worked as a visiting researcher at the University of Texas at Dallas in the United States from 2009 to 2010. At present, he is the president of the school of geography and science, the director of the Key Laboratory of the Ministry of education of the virtual geographical environment, the executive director of the Collaborative Innovation Center for the development and utilization of geographic information resources in Jiangsu Province, and the deputy director of the academic evaluation committee of the Chinese geographical society. He published more than 50 journal papers, and won the first prize of natural science of the Ministry of education, the first prize of national geographic information technology progress.

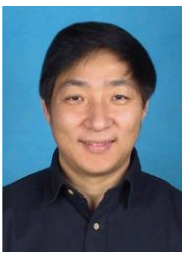

**A-Xing Zhu** received the Ph.D. degree from University of Toronto, is now a Manasse Chair Professor in the Geography Department of University of Wisconsin-Madison. His research interests are GIS/RS, artificial intelligence, fuzzy logic, and watershed system modeling and scenario analysis for best management practices, intelligent geocomputing. He is the Editor-in-Chief of "Annals of GIS".